\documentclass[fleqn,10pt]{SelfArx} % Document font size and equations flushed left

\usepackage[english]{babel} % Specify a different language here - english by default

\usepackage{lipsum} % Required to insert dummy text. To be removed otherwise
\usepackage{tikz}
\usetikzlibrary{fit,positioning,calc}

\definecolor{color1}{RGB}{0,0,90} % Color of the article title and sections
\definecolor{color2}{RGB}{0,20,20} % Color of the boxes behind the abstract and headings

\usepackage{hyperref} % Required for hyperlinks
\JournalInfo{Journal, Vol. XXI, No. 1, 1-5, 2013} % Journal information
\Archive{Additional note} % Additional notes (e.g. copyright, DOI, review/research article)

\PaperTitle{Unifying inference on brain network variations in neurological diseases: The Alzheimer's case} % Article title

\Authors{Daniele Durante\textsuperscript{1}, Madelaine Daianu\textsuperscript{2}, Neda Jahanshad\textsuperscript{2}, Paul M. Thompson\textsuperscript{2}, David B. Dunson\textsuperscript{3}} % Authors
\affiliation{\textsuperscript{1}\textit{University of Padova, Padova,Italy}} % Author affiliation
\affiliation{\textsuperscript{2}\textit{University of Southern California, Los Angeles, USA}} % Author affiliation
\affiliation{\textsuperscript{3}\textit{Duke University, Durham, USA}} 
 % Keywords - if you don't want any simply remove all the text between the curly brackets
\newcommand{\keywordname}{Keywords} % Defines the keywords heading name

\Abstract{There is growing interest in understanding how the structural interconnections among brain regions change with the occurrence of neurological diseases.  Diffusion weighted MRI imaging has allowed researchers to non-invasively estimate a network of structural cortical connections made by white matter tracts, but current statistical methods for relating such networks to the presence or absence of a disease cannot exploit this rich network information.  Standard practice considers each edge independently or summarizes the network with a few simple features.  We enable dramatic gains in biological insight via a novel unifying methodology for inference on brain network variations associated to the occurrence of neurological diseases.  The key of this approach is to define a probabilistic generative mechanism directly on the space of network configurations via dependent mixtures of low-rank factorizations, which efficiently exploit network information and allow the probability mass function for the brain network-valued random variable to vary flexibly across the group of patients characterized by a specific neurological disease and the one comprising age-matched cognitively healthy individuals.}

\begin{document}

\flushbottom % Makes all text pages the same height

\maketitle % Print the title and abstract box

\tableofcontents % Print the contents section

\thispagestyle{empty} % Removes page numbering from the first page

%----------------------------------------------------------------------------------------
%	ARTICLE CONTENTS
%----------------------------------------------------------------------------------------

\section{Introduction} % The \section*{} command stops section numbering

There is fundamental interest in understanding the relationship between brain connectivity structure and neurodegenerative disorders, such as Alzheimer's \cite{ASH}, Parkinson's  \cite{EID} and other dementias \cite{BUD}. These diseases are mainly found in aged populations and affect the normal functions of the central and peripheral nervous system causing, among others, muscle weakness, loss of coordination and cognitive impairment.

Alarming prevalence projections of dementia cases by the World Health Organization in 2006 \cite{WHO}, and the rapid development of brain imaging technologies in recent years, have stimulated intensive research aimed at understanding how the brain structure is compromised with specific neurological diseases. This is key to improving  diagnosis as well as providing increasingly targeted therapies.

Most of the related literature has focused on the modular paradigm, which considers brain regions as specialized actors in specific cognitive functions \cite{FOD}. Under this paradigm, inference focuses on multivariate data $\psi_i=(\psi_{i1}, \ldots, \psi_{iV})^T$, where $\psi_{iv}$ is the activity level in region $v$ $(v=1, \ldots, V)$ for individual $i$ $(i=1,\ldots,n)$, often measured via functional magnetic resonance imaging (fMRI).  Let $y_i$ denote an indicator of whether individual $i$ is in the control group ($y_i=0$) or has a specific disorder of interest ($y_i=1$).  Inference then proceeds by testing for global and local variations in the brain activity vector across these two groups.

In contrast to functional correlations, diffusion weighted magnetic resonance imaging (dw-MRI) can provide information on the structural connection network in the brain by approximating the diffusion of water molecules across tissues.  Water diffuses equally in all directions in grey matter, but is constrained and does not similarly diffuse across white matter fibers.  This allows for the reconstruction of white matter tracts using tractography methods for dw-MRI \cite{CRA} and mapping the patterns of connections formed from the white matter fiber bundles.  Using this information, we can obtain measurements of the brain's network of cortical connectivity by seeing if certain anatomical regions are connected by white matter. 
\begin{figure}
\centerline{\includegraphics[width=.45\textwidth]{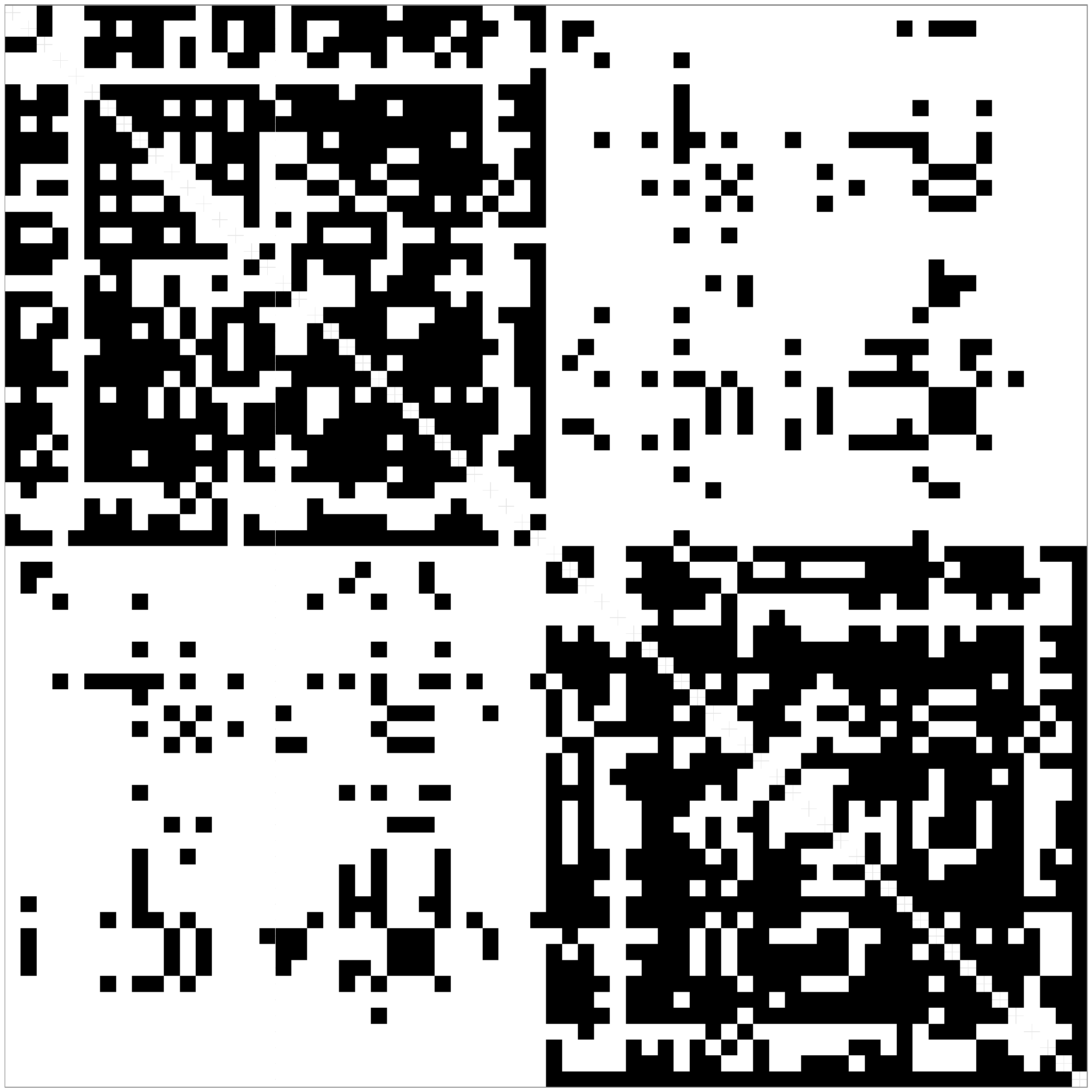}}
\caption{Adjacency matrix $A_i$ representing the brain network of the $i$-th subject. Black corresponds to edges, white to non-edges. In our Alzheimer's study, brain anatomical regions are defined by the Desikan atlas \cite{DES} for a total of $V=68$ nodes equally divided between left and right hemisphere.}\label{F1}
\end{figure}

In this article, we focus on brain network data corresponding to a $V \times V$ symmetric binary adjacency matrix $A_i$ for individual $i$ having elements $A_{i[vu]}=A_{i[uv]}=1$ if there is one or more white matter tracts connecting brain regions $v$ ($v=2,\ldots,V$) and $u$ ($u=1, \ldots, v-1$) in individual $i$ $(i=1,\ldots,n)$ and  $A_{i[vu]}=A_{i[uv]}=0$ otherwise.  Refer to Fig. 1 for an illustrative example.  Such connectome data provide substantial insights into the sources of complex cognitive processes relative to brain activity data analyzed in the modular paradigm  \cite{FUS}. Cognitive network theory views the studying of structural connectivity networks as a key to learning the complexity of information processing mechanisms and understanding the effect of neurological disorders on these cognitive processes  \cite{BRE}.

Much of today's literature is focused on analytic methods for understanding localized brain activity data $\psi_i$, yet methodologies for analyzing brain network data $A_i$ is still in its infancy.  Our main aim is to develop techniques to assess whether and how a network-valued random variable generating structural brain networks $A_i$ $(i=1, \ldots, n)$ varies across diagnostic groups.  In particular, it is of interest to test for global variation in the overall brain network structure across groups, while identifying specific local variations to understand if and which brain connections are compromised by a specific neurological disease of interest.

There has been an increasing attention in the literature toward methods for addressing these aims; see \cite{STA} and the references cited therein for an overview.  Common practice proceeds by reducing $A_i$ to summary measures $\theta_i=(\theta_{i1},\ldots,\theta_{ip})^T$ $(i=1, \ldots,n)$ and then applying standard multivariate analyses such as MANOVA (see e.g. \cite{KRZ}) to assess whether these network measures change with the occurrence of a disease.  Summary statistics are commonly chosen to represent global network characteristics, such as the average path length, network density, transitivity and k-core \cite{RUB}. These measures provide a useful simplification of a complex problem, but cannot characterize the entire network structure and hence may fail to detect important relationships between brain network and neurological disorders, leading to inconsistent results in the literature; see \cite{ARD} for a review of inconsistencies when relating brain networks to creative reasoning.  

An alternative class of methods avoid reducing brain networks to summary statistics by performing separate and independent tests to assess which  brain connections are compromised by a neurological disease. As there are $V(V-1)/2$ edges in the $V$ brain regions under study, with $V=68$ using the Desikan atlas \cite{DES}, the number of tests is  substantial and requires  adjustments of the significance threshold to control for multiplicity \cite{GEN}.  Such mass-univariate approaches do not exploit network information, leading to low power \cite{FOR}, and substantially underestimating the number of connections compromised by a specific neurological disorder.  Recent proposals try to gain power by replacing the false discovery rate (FDR) control \cite{BEN}, with thresholding procedures that account for the network structure \cite{ZAL}.  However, such approaches require careful interpretation, while being highly computationally intensive and complex.

We propose a fundamentally new approach based on defining a generative probabilistic model for the brain network data.  In particular, the probability mass function (pmf) for the network-valued random variable is assigned a mixture model, allocating individuals to subpopulations in terms of their brain network structure.  Within a subpopulation, the edge probabilities are related to a latent similarity measure via a logistic mapping.  The similarity matrix is then factorized as the sum of a common component and a subpopulation-specific one that arises from embedding the brain regions in a low dimensional latent space that accounts for the network structure.  Using this flexible mixture model as scaffolding, we develop an efficient testing method by allowing the mixture weights to vary across case and control groups.  This induces a highly efficient Bayesian testing procedure that adjusts automatically for multiple comparisons in drawing inferences on brain structural differences across groups.

In the rest of the paper, we describe the model formulation, providing insights on theoretical properties, estimation procedures and inference techniques. An application to assess variations in the brain architecture with Alzheimer's disease illustrates the benefits of the proposed procedure, while providing novel insights into the impact of the disease on brain structural connections.

\section{Dependent Mixture of Low-rank Factorizations}
A network-valued random variable $\mathcal{A}$ generating symmetric $V \times V$ binary adjacency matrices can be uniquely characterized by its lower triangular matrix elements $\mathcal{L}(\mathcal{A})=(\mathcal{A}_{21}, \mathcal{A}_{31}, \ldots, \mathcal{A}_{V1}, \mathcal{A}_{32}, \ldots,  \mathcal{A}_{V2}, \ldots, \mathcal{A}_{V(V-1)})^{T} $, as $\mathcal{A}_{[vu]}=\mathcal{A}_{[uv]}$ for every $v=2, \ldots, V$, $u=1, \ldots, v-1$ and the diagonal elements are not of interest for inference. Since there are finitely many network configurations, $\mathcal{L}(\mathcal{A})$ can be seen as a categorical variable with each category representing one of the possible network configurations $a \in \mathbb{A}_V= \{0,1\}^{V(V-1)/2}$. As there are $2^{V(V-1)/2}$ possible undirected network configurations among $V$ nodes, $2^{V(V-1)/2}-1$ parameters are required to characterize the pmf $p_{\mathcal{L}(\mathcal{A})}(a)=\mbox{Pr}\{\mathcal{L}(\mathcal{A})=a\}$, $a \in \mathbb{A}_V$ under the the usual restriction $\sum_{a \in  \mathbb{A}_V} p_{\mathcal{L}(\mathcal{A})}(a)=1$.

This number becomes intractable and massively larger than the sample size $n$ even in using coarse brain regions. In the motivating Alzheimer's disease study, there are $V=68$ brain regions.  This implies that, in the absence of constraints, we need to estimtae  $2^{68(68-1)/2}-1=2^{2278}-1$ free parameters characterizing $p_{\mathcal{L}(\mathcal{A})}$.  Clearly no studies will ever have this many subjects, and hence it is necessary to  reduce dimensionality and make the problem tractable.  However, in reducing dimension, it is important to maintain flexibility in characterizing the structure underlying brain networks. 

We propose to reduce dimensionality, while maintaining flexibility in characterizing $p_{\mathcal{L}(\mathcal{A})}$, by using a hierarchical latent space representation. The idea is to assign each brain region a coordinate in a lower dimensional Euclidean space; such models have been used effectively in social network contexts \cite{HOFF}.  Our contribution is the generalization to a latent space random effects model which defines the population distribution of network-valued data, while characterizing individual  differences in the architecture of interconnections in the brain.  The random effects are modeled as a mixture of low-rank factorizations.  
This induces clustering of individuals in terms of their brain structure, and facilitates inferences on differences in case and controls by allowing the mixing weights to vary across these groups.  We first describe the low-rank factorization structure, which represents the key building block to reduce dimensionality and borrow network information.
\begin{figure}
\begin{center}
\centerline{\includegraphics[height=7.7cm, width=9.2cm]{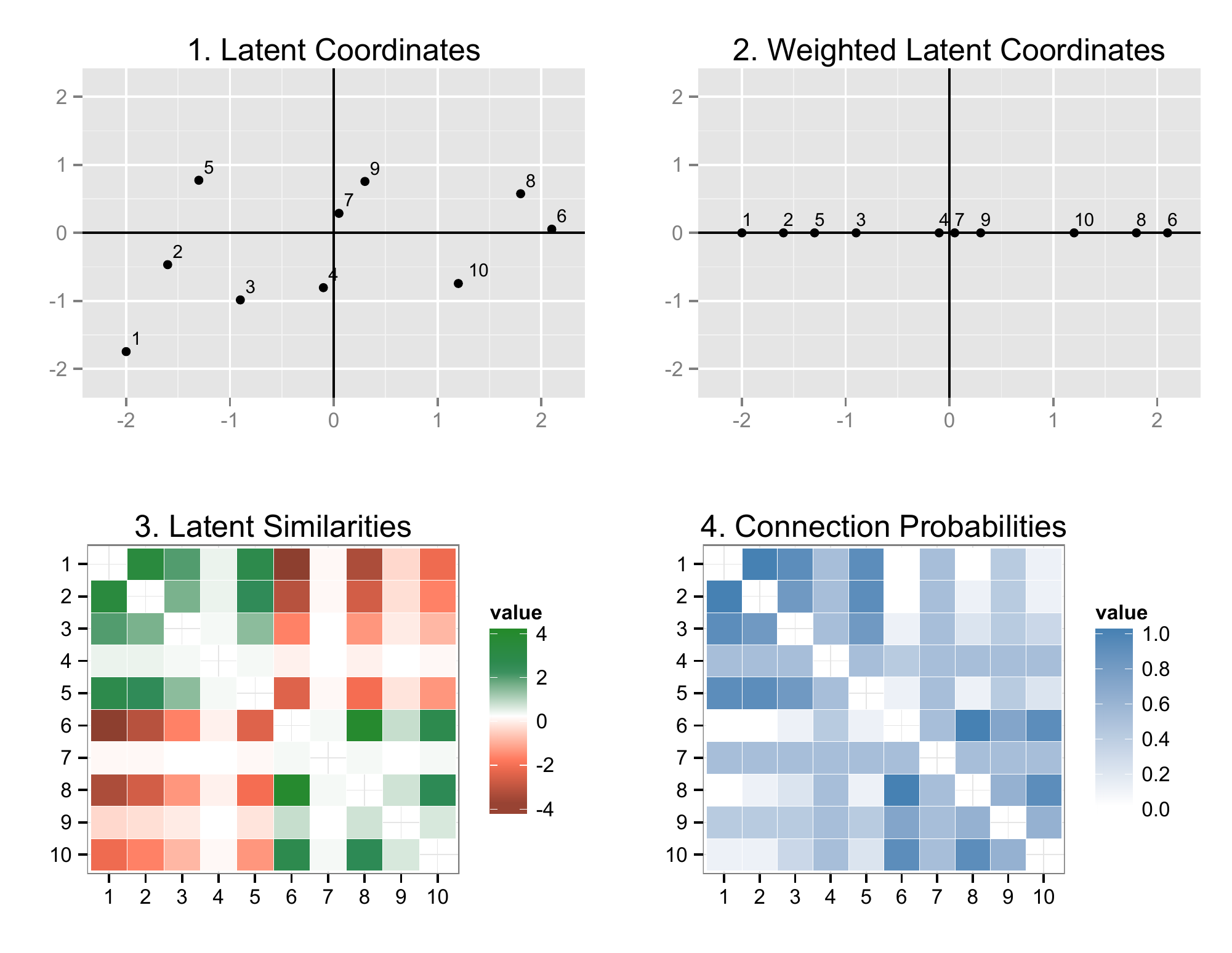}}
\caption{Steps to construct the connection probabilities via low-rank factorization.}\label{F2}
\end{center}
\end{figure}
\subsection{Low-rank factorization}
Letting $\pi=(\pi_1, \ldots, \pi_{V(V-1)/2})^T \in (0,1)^{V(V-1)/2}$ denote the vector of probabilities of edges between each pair of brain regions, our probabilistic low-rank factorization generates brain networks in two main steps displayed in Fig. 2 and Fig. 3. The first constructs the edge probability vector $\pi$ exploiting network information as follows:

\begin{enumerate}
\item{Embed each region $v$ in an $R$-dimensional Euclidean space (in Fig. 2, $R=2$), with $X$ a $V \times R$ matrix of latent coordinates; $X_{vr}$ is the $r$th coordinate for brain region $v$.}
\item{The $r$th column of $X$ corresponds to the coordinates for the different brain regions in dimension $r$.  These latent coordinates are sampled from a standard normal distribution, and then multiplied by $\sqrt{\lambda}_r$, with $\lambda_r$ measuring the overall importance of the $r$th dimension of the latent space on brain network structure. In Fig. 2 $\lambda_1=1$ while $\lambda_2=0$, meaning that the second coordinate doesn't play any role in generating the network.}
\item{Focusing on the $l$th pair of brain regions, corresponding to nodes $v$ and $u$, $v>u$, construct the similarity measure $S_l$ via dot product of their weighted latent coordinate vectors $S_l=\sum_{r=1}^{R}\lambda_r X_{vr} X_{ur}$,  for each $l=1, \ldots, V(V-1)/2$.}
\item{Define $\pi_l$ by mapping each similarity measure $S_l \in \Re$ into the probability space via the one-to-one continuous increasing logistic mapping, so that $\pi_l=\{1+\exp(-S_l)\}^{-1}$ for each $l=1, \ldots, V(V-1)/2$.}
\end{enumerate}

The second step generates networks $\mathcal{L}(A_i)$ by sampling their edges $\mathcal{L}(A_i)_l$ $(l=1,\ldots, V(V-1)/2)$ from conditionally independent Bernoulli random variables given  connection probabilities $\pi_l=\mbox{Pr}\{\mathcal{L}(\mathcal{A}_i)_l=1 \} \in (0,1)$. Although the mechanism generates networks with conditionally independent edges given $\pi$, the shared dependence on a common set of node-specific latent coordinates induced by the dot product representation of $S=(S_1, \ldots, S_{V(V-1)/2})^T \in \Re^{V(V-1)/2}$ can define arbitrarily rich dependence structures. This is illustrated in Fig. 3, where the networks preserve a two-block structure induced by factorization in Fig. 2. The low-rank factorization allows dimensionality reduction from $V(V-1)/2$ edge probabilities to $V \times R$ latent coordinates and $R$ weights.  The formulation facilitates adaptive collapsing on lower dimensional models by shrinking the weights $\lambda_r$ towards $0$ as $r$ increases. 

\begin{figure}
\begin{center}
\centerline{\includegraphics[width=.45\textwidth]{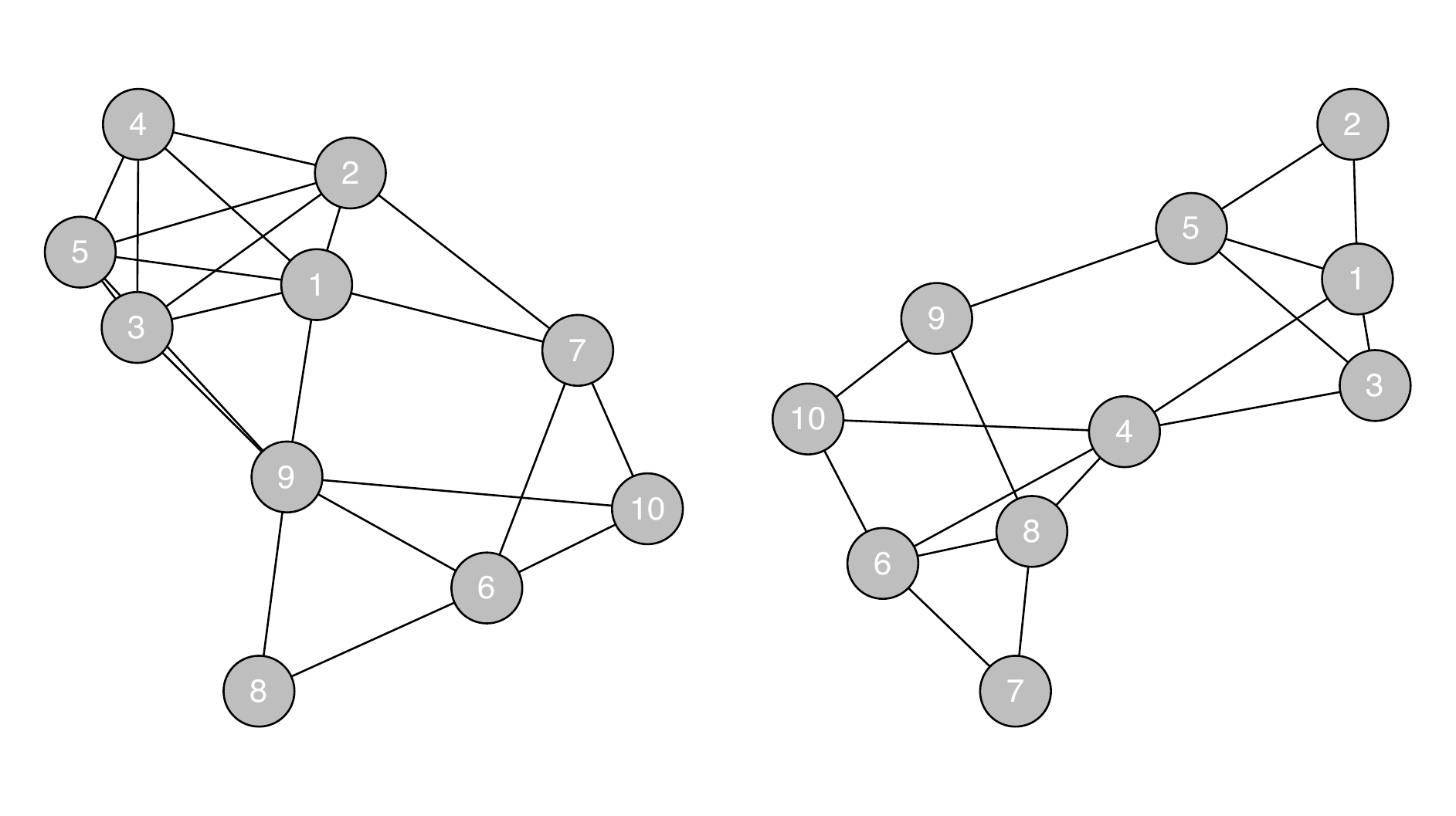}}
\caption{Two networks generated from the low-rank factorization mechanism with connections probabilities from Figure 2.}\label{F3}
\end{center}
\end{figure}

\subsection{Mixture of low-rank factorizations}
Under the proposed model, the probability assigned to configuration  $a \in  \mathbb{A}_V$ is
\begin{eqnarray}
p_{\mathcal{L}(\mathcal{A})}(a)&=& \prod_{l=1}^{V(V-1)/2} \mbox{Pr}\{\mathcal{L}(\mathcal{A})_l=a_l\}, \nonumber\\
&=&\prod_{l=1}^{V(V-1)/2} \mbox{Bern}(a_l;\pi_l),
\label{eq:1}
\end{eqnarray}
where $ \mbox{Bern}(a_l;\pi_l)=\pi_l^{a_l}(1-\pi_l)^{1-a_l}$ and $\pi$ is generated according to Fig. 2. Although providing a key building block to reduce dimensionality and include network information, the low-rank factorization  is insufficiently flexible in assuming independence across pairs of brain regions in the occurrence of connections conditionally on $\pi$. To improve flexibility, we generalize the model to mix together $H$ subpopulations.

Let $G_i \in \{1,\ldots, H\}$ index the subpopulation that the $i$th brain network $A_i$ is generated from, with $p_G(h)=\mbox{Pr}(G=h)=\nu_h$ the probability of being drawn from that subpopulation, and 
$\pi^{(h)}$ the connections probability vector specific to that subpopulation.  Then, we have the following generative process: 
\begin{enumerate}
\item{Allocate the $i$th individual to a subpopulation by sampling $G_i$ according to $p_G$.}
\item{Given $G_i=h$ and the corresponding $\pi^{(h)}$, generate $A_i$ by sampling its edges $\mathcal{L}(A_i)_l$ $(l=1,\ldots, V(V-1)/2)$ from conditionally independent Bernoulli random variables given the connection probabilities specific to subpopulation $h$, $\pi^{(h)}_l=\mbox{Pr}\{\mathcal{L}(\mathcal{A}_i)_l=1 \mid G_i=h \} \in (0,1)$.}
\end{enumerate}
By marginalizing out the subpopulation indicator $G_i$, this hierarchical probabilistic generative process leads to a mixture representation for $p_{\mathcal{L}(\mathcal{A})}$ of the form
\begin{eqnarray}
p_{\mathcal{L}(\mathcal{A})}(a)= \sum_{h=1}^H \nu_h\prod_{l=1}^{V(V-1)/2} \mbox{Bern}\{a_l;\pi^{(h)}_l\}.
\label{eq:2}
\end{eqnarray}
Equation \eqref{eq:2} is much more flexible than \eqref{eq:1}.

Statistical properties of representation \eqref{eq:2} have been studied in \cite{DUR1}, showing the full flexibility of the mixture of low-rank factorizations in representing every possible pmf $p_{\mathcal{L}(\mathcal{A})}$. They additionally slightly modify the steps generating the class-specific edge probability vectors $\pi^{(h)}=\{1+\exp(-S^{(h)})\}^{-1}$ $(h=1, \ldots, H)$ to allow inference on shared versus class-specific components of variability in the brain connectivity structure. In particular, instead of defining the entries in the latent similarity vector for the $h$th subpopulation $S^{(h)}$ as $S^{(h)}_l=\sum_{r=1}^{R}\lambda^{(h)}_r X^{(h)}_{vr} X^{(h)}_{ur}$, they let
\begin{eqnarray}
\mbox{logit}(\pi^{(h)}_l)=S^{(h)}_l=Z_l+D_l^{(h)},\  \ D_l^{(h)}= \sum_{r=1}^{R}\lambda^{(h)}_r X^{(h)}_{vr} X^{(h)}_{ur},
\label{eq:3}
\end{eqnarray}
for each $l=1, \ldots, V(V-1)/2$, with $Z_l \in \Re$ a latent similarity measure for the $l$th pair of regions common to all individuals and $D_l^{(h)} \in \Re$ a deviation specific to subpopulation $h$. The common similarity $Z_l$ in \eqref{eq:3} is left unstructured as it can be estimated borrowing information across all individuals, while $D_l^{(h)}$ is constructed according to steps $1-3$ of the low-rank factorization mechanism for each $h=1, \ldots, H$ exploiting the network structure to cope with less information in the data about class-specific deviations. 

\subsection{Inferences on differences across groups}
When the focus is on inference on changes in brain networks with the occurrence of  a disease, the model described above needs to be generalized.  This can be accomplished by defining a joint pmf for the random variable $\{\mathcal{Y}, \mathcal{L}(\mathcal{A})\}$ generating data 
 $\{y_i,\mathcal{L}(A_i)\}$ $(i = 1,\ldots,n)$, which allows testing of the global association between $\mathcal{Y}$ and $ \mathcal{L}(\mathcal{A})$ as well as local dependence between $\mathcal{Y}$ and each edge $\mathcal{L}(\mathcal{A})_l$. 
 
Let $p_{\mathcal{Y}, \mathcal{L}(\mathcal{A})}$ denote the joint pmf for the random variable $\{\mathcal{Y}, \mathcal{L}(\mathcal{A})\}$, with $p_{\mathcal{Y}, \mathcal{L}(\mathcal{A})}(y,{a})=\mbox{Pr}\{\mathcal{Y}=y, \mathcal{L}(\mathcal{A})={a}\}$, $y \in \{0,1\}$ and ${a} \in \mathbb{A}_V$ a network configuration. We first characterize  the joint pmf $p_{\mathcal{Y}, \mathcal{L}(\mathcal{A})}$ as the product of the marginal ${p}_{\mathcal{Y}}$ for the group variable and the conditional pmfs ${p}_{ \mathcal{L}(\mathcal{A}) \mid y}$, $y \in \{0,1\}$, of the brain network-valued random variable given the presence or absence of the neurological disorder, obtaining
\begin{eqnarray}
p_{\mathcal{Y}, \mathcal{L}(\mathcal{A})}(y,{a})&=& p_{\mathcal{Y}}(y){p}_{ \mathcal{L}(\mathcal{A}) \mid y}(a)\nonumber\\
&=&\mbox{Pr}(\mathcal{Y}=y)\mbox{Pr}\{\mathcal{L}(\mathcal{A})=a \mid \mathcal{Y}=y\}. 
\label{eq:4}
\end{eqnarray}
Although we treat $\mathcal{Y}$ as a random variable through a prospective likelihood, the method we propose is valid for inference on differences across groups in brain network structure also for case control studies that sample under a retrospective design. 

The conditional pmf for the brain network given the presence or absence of the disease, ${p}_{ \mathcal{L}(\mathcal{A}) \mid y}$, $y \in \{0,1\}$, is characterized via a slight modification of the mixture of low-rank network factorization proposed in the previous section.  In particular, we simply allow the proportions of individuals in each subpopulation to vary across groups by letting $p_{G\mid y}(h)=\mbox{Pr}(G=h \mid \mathcal{Y}=y)=\nu_{hy}$, while holding the subpopulation specific edge probability vectors ${\pi}^{(h)}$ fixed.  A graphical representation of the final model is provided in Fig. 4. Replacing $\mbox{Pr}(G=h)=\nu_h$ with $\mbox{Pr}(G=h \mid \mathcal{Y}=y)=\nu_{hy}$ $(h=1, \ldots, H)$ in equation \eqref{eq:2}, leads to the dependent mixture representation
\begin{eqnarray}
{p}_{ \mathcal{L}(\mathcal{A}) \mid y}(a)&=& \sum_{h=1}^H \nu_{hy}\prod_{l=1}^{V(V-1)/2} \mbox{Bern}\{a_l;\pi^{(h)}_l\}.
\label{eq:5}
\end{eqnarray}
for control ($y=0$) and case ($y=1$) group. According to theoretical properties in \cite{DUR2},  factorizations $\eqref{eq:4}-\eqref{eq:5}$, with each $\pi^{(h)}$ constructed as in  equation \eqref{eq:3} are fully flexible in characterizing any joint pmf $p_{\mathcal{Y}, \mathcal{L}(\mathcal{A})}$, while reducing the dimensionality in requiring estimation of $H$ low-rank factorization mechanisms, rather than $2^{V(V-1)/2}-1$ configuration probabilities for each group $y$. 

Under this formulation, testing of the null hypothesis of no differences in the distribution of brain networks between the case and control groups versus the alternative of some difference can be mathematically expressed as 
\begin{eqnarray}
H_0: p_{G \mid 0}=p_{G \mid 1}=p_{G} \ \ \mbox{versus} \ \ H_1: p_{G \mid 0}\neq p_{G \mid 1}.
\label{eq:6}
\end{eqnarray}
Local inferences on whether individual edge probabilities vary with the presence of a neurological disorder can instead be based on Cramer's V \cite{DUN}
\begin{eqnarray}
\rho^2_{l}=\sum_{y=0}^1 p_{\mathcal{Y}}(y) \sum_{a_l=0}^1\frac{\left\{p_{\mathcal{L}(\mathcal{A})_l \mid y}(a_l)-p_{\mathcal{L}(\mathcal{A})_l}(a_l)\right\}^2}{p_{\mathcal{L}(\mathcal{A})_l}(a_l)},
\label{ro}
\end{eqnarray}
 for $l=1, \ldots, V(V-1)/2$, where $p_{\mathcal{Y},\mathcal{L}(\mathcal{A})_l}(y,a_l)=\mbox{Pr}\{\mathcal{Y}=y, \mathcal{L}(\mathcal{A})_l=a_l\}$ and $p_{\mathcal{L}(\mathcal{A})_l}(a_l)=\mbox{Pr}\{\mathcal{L}(\mathcal{A})_l=a_l\}$. The coefficient $\rho_{l}$ has values in the interval $[0,1]$, with $\rho_{l} = 0$ meaning that ${p}_{\mathcal{Y},\mathcal{L}(\mathcal{A})_l} = {p}_{\mathcal{Y}}{p}_{\mathcal{L}(\mathcal{A})_l}$, so that the probability of a connection  between the $l$th pair of brain regions is the same for cases and controls.
\begin{figure}[t]
\centering
\begin{tikzpicture}
\tikzstyle{main}=[circle, minimum size = 10mm, thick, draw =black!80, node distance = 16mm]
\tikzstyle{connect}=[-latex, thick]
\tikzstyle{box}=[rectangle, draw=black!100]
  \node[main, fill = white!100] (theta) {$\nu_y$ };
  \node[main] (z) [right=of theta] {$G_i$};
  \node[main] (beta) [above=of z] {$S^{(h)}$ };
    \node[main,fill = black!10] (d) [below=of z] {$y_i$ };
        \node[main] (p) [left=of d] {$p_{\mathcal{Y}}$ };
  \node[main, fill = black!10] (w) [right=of z] {$A_i$ };
    \node[main] (pi) [above=of w] {$\pi^{(h)}$ };
      \node[main] (x) [above= of {$(pi)!0.5!(beta)$}]  {$X^{(h)}$};
            \node[main] (lam) [left= of x]  {$\lambda^{(h)}$};
             \node[main] (Z) [left= 2.5cm of beta]  {$Z$};
  \path        (theta) edge [connect] (z)
        (z) edge [connect] (w)
        (beta) edge [connect] (pi)
        (pi) edge [connect] (w)
        (d) edge [connect] (z)
        (x) edge [connect] (beta)
                (Z) edge [connect] (beta)
        (lam) edge [connect] (beta)
                (p) edge [connect] (d);
  %\node[rectangle, inner sep=0mm, fit= (z) (w),label=below right:n, xshift=13mm] {};
 % \node[rectangle, inner sep=4.4mm,draw=black!100, fit= (z) (w)] {};
 % \node[rectangle, inner sep=4.6mm, fit= (z) (w),label=below right:M, xshift=12.5mm] {};
  %\node[rectangle, inner sep=9mm, draw=black!100, fit = (theta) (z) (w)] {};
  \node[rectangle, inner sep=6.5mm, draw=black!100, fit = (theta)] {};
  \node[rectangle, inner sep=0mm, fit= (theta)] {};
  \node[rectangle, inner sep=7mm, draw=black!100, fit = (d) (z) (w)] {};
    \node[rectangle, inner sep=7mm, draw=black!100, fit = (pi) (beta) (x) (lam)] {};
  \node[] at (0.2,-0.89) {$y \in \{0,1\}$};
    \node[] at (5.2,-3.5) {$i=1, \ldots, n$};
        \node[] at (1.2,1.7) {$h=1, \ldots, H$};
    %\node[rectangle, inner sep=0mm, fit= (d) (z) (w),label={[shift={(1.0,0.3)}]Label}] {};
\end{tikzpicture}
\caption{Graphical representation of the dependent mixture of low-rank factorization model.}\label{F_dep}
\end{figure}
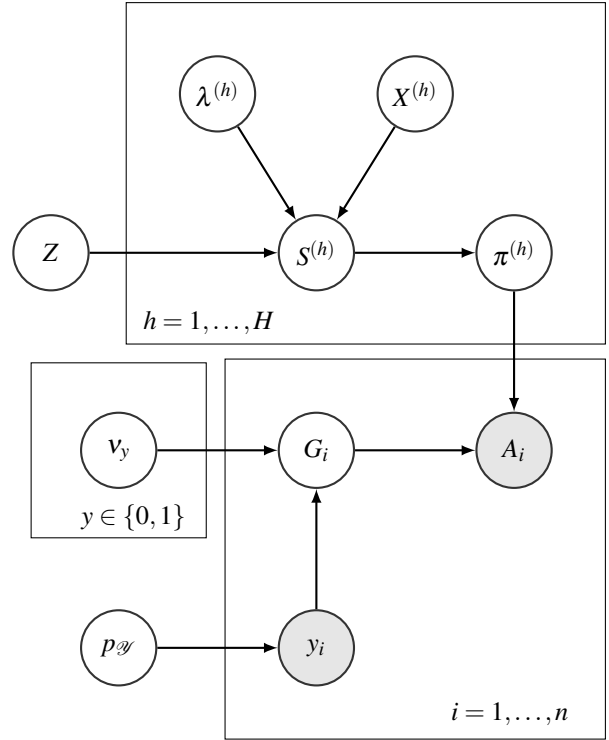

\subsection{Bayesian inference}
Inference for the model in Fig. 4 is performed under a Bayesian paradigm. Bayesian inference characterizes uncertainty and facilitates adaptive choice of $R$ and $H$ through carefully specified priors.  Hence, the focus is on the posterior distribution
\begin{eqnarray}
\Pi\{{p}_{\mathcal{Y},\mathcal{L}(\mathcal{A})}  \mid y_1, \mathcal{L}(A_1), \ldots, y_n,  \mathcal{L}(A_n)\}.
\label{eq:7}
\end{eqnarray}
As the joint ${p}_{\mathcal{Y},\mathcal{L}(\mathcal{A})}$ is defined via equations $\eqref{eq:3}-\eqref{eq:5}$, the posterior distribution in equation \eqref{eq:7} can be easily obtained as a function of the posteriors for parameters in $\eqref{eq:3}-\eqref{eq:5}$. Inference proceeds by updating the independent priors for the quantities ${p}_{\mathcal{Y}}\sim\Pi_{y}$, ${Z}=(Z_1, \ldots, Z_{V(V-1)/2}) \sim \Pi_Z$, ${X}^{(h)} \sim \Pi_{X}$, ${\lambda}^{(h)}=(\lambda^{(h)}_{1}, \ldots, \lambda^{(h)}_{R}) \sim \Pi_{\lambda}$,  $h=1, \ldots, H$ and ${\nu}_{y}=(\nu_{1y}, \ldots, \nu_{Hy})\sim \Pi_{\nu}$, $y \in \{0,1\}$ given brain network data and case-control status, to obtain posterior samples via MCMC methods. We compute \eqref{eq:7} as a function of these posteriors via  $\eqref{eq:3}-\eqref{eq:5}$.

Prior distributions are carefully defined to induce a prior $\Pi$ on the joint pmf ${p}_{\mathcal{Y},\mathcal{L}(\mathcal{A})}$ with simple posterior computation procedures, allowance for testing and flexibility. These aims are accomplished by letting $p_{\mathcal{Y}}(0)=1-p_{\mathcal{Y}}(1) \sim \mbox{Beta}(a_0,a_1)$, while  choosing Normal priors for the entries in $Z$ and standard Gaussians for the elements in $X^{(h)}$. To learn the dimensions of the latent spaces and penalize high dimensional representations, a multiplicative inverse gamma is defined for ${\lambda}^{(h)}$. This choice for $\Pi_{\lambda}$ favors shrinkage effects with elements in ${\lambda}^{(h)}$ stochastically decreasing towards $0$ as $r$ increases; see \cite{DUR1} for details.

Prior $\Pi_{\nu}$ is instead defined to incorporate global hypothesis testing in equation \eqref{eq:6}, while allowing automatic deletion of redundant classes. This is accomplished by introducing an hypothesis indicator $T \in \{0,1 \}$, with $T=0$ for $H_0$ and $T=1$ for $H_1$. If $T=1$ the latent class assignment mechanism is different between case and controls and two independent Dirichlet priors priors are assigned to $\nu_0$ and $\nu_1$, respectively. If $T=0$, we set $\nu_{0}=\nu_{1}=u \sim   \mbox{Dir}(1/H, \ldots, 1/H)$ as $p_G$ is the same in the two groups, under $H_0$. Hence in assessing evidence in favor of the alternative, one  can rely on the posterior probability, $\mbox{Pr}\{H_1 \mid y_1, \mathcal{L}(A_1), \ldots, y_n,  \mathcal{L}(A_n)\}=\mbox{Pr}\{T=1 \mid  y_1, \mathcal{L}(A_1), \ldots, y_n,  \mathcal{L}(A_n)\}$. This quantity can be easily computed in the MCMC algorithm for posterior computation; see \cite{DUR2} for details. Beside this benefit for global hypothesis testing, deletion of redundant classes can also be easily accomplished by choosing small values for the hyperparameters in the Dirichlet priors \cite{ROU}.

Local testing of edge probability differences between case and control groups proceeds via interval nulls $H_{0l}: \rho_l \leq \epsilon$ versus $H_{1l}: \rho_l > \epsilon$. As noted in \cite{BER} the small interval hypothesis $H_{0l}: \rho_l \leq \epsilon$ can be realistically approximated by $H_{0l}: \rho_l =0$, moreover this choice facilitates the computation of  $\mbox{Pr}\{H_{1l} \mid y_1, \mathcal{L}(A_1), \ldots, y_n,  \mathcal{L}(A_n)\}=1-\mbox{Pr}\{H_{0l} \mid  y_1, \mathcal{L}(A_1), \ldots, y_n,  \mathcal{L}(A_n)\}$ as the proportion of MCMC samples in which $\rho_l >\epsilon$.

Simulation studies in \cite{DUR2} highlight the good performance of the proposed methodology in accurately estimating the joint pmf ${p}_{\mathcal{Y},\mathcal{L}(\mathcal{A})}$ and in accurately identifying differences between case and controls in the brain network structure, while  identifying local variations in each edge probability. 
Across multiple scenarios, the proposed local testing procedure has a Type I error of $0.00044$, Type II of $0.0587$ and FDR of $0.0023$. Independent screening via separate two-sided Fisher's exact tests (see e.g. \cite{AGR}) with FDR control has Type I error of $0.0036$, Type II of $0.5983$ and FDR of $0.0387$. These improvements are also exhibited in global testing with our procedure having both Type I and II errors of $0.01$. In contrast MANOVA test on network features has Type I error of $0.10$ and Type II error of $0.87$. 

\section{Application to Alzheimer's Disease}
According to the Centers for Disease Control and Prevention (CDC), Alzheimer's disease (AD) is the most common form of dementia and the sixth leading cause of death in the United States.  Unlike cancer and heart disease death rates, which are expected to decline, the growth of elderly population in the age range most commonly affected by dementia is leading to an increase of the death rates due to AD \cite{JAM}. This has strongly motivated intensive research aimed at finding the sources of AD in the human brain to develop increasingly refined diagnosis and prognosis procedures and improve therapy.

Current understanding of variations in brain behavior across AD is mostly available via early neuropathological studies (e.g. \cite{HOP}),  and contributions analyzing joint or local changes in the activity of each region under the modular paradigm (e.g. \cite{THO}). More recent proposals shift increasingly away from the above approach towards studying brain activity networks via changes of the covariance in activity across brain regions for AD and controls (e.g. \cite{BOK}). However, functional connectivity matrices estimated from fMRI data do not reflect the underlying axonal pathways that can give rise to changes in function, and often require caution in interpreting the results \cite{BRE}. This has motivated an increasing interest in structural connectivity matrices estimated from diffusion scans. Early studies on these data proceed by assessing variations of global brain network measures or region-specific connectivity statistics across AD and controls (e.g. \cite{DAI}). As previously noted, these methods may fail in flexibly characterizing the richness of the brain network structure, leading to inconsistent results. 

To address these issues, we apply our methodology to brain networks derived from the Alzheimer's Disease Neuroimaging Initiative (ADNI), providing consistent new insights which contribute to solving the ongoing mystery behind the mechanisms of AD in the human brain. Data $(y_i,A_i)$ are available for $n=92$ individuals, with $50$  in the control group $(y_i=0)$ and $42$ age-matched patients with AD $(y_i=1)$. Each adjacency matrix $A_i$ represents the brain network of the $i$-th individual, measuring the structural connectivity among $V=68$ lateralized brain regions defined by the Desikan atlas \cite{DES} as part of FreeSurfer \cite{FISH}. In particular $A_i$ has elements $A_{i[vu]}=A_{i[uv]}=1$ if at least one white matter tract connects brain regions $v$ and $u$ in individual $i$ and  $A_{i[vu]}=A_{i[uv]}=0$, otherwise. These structural networks, also known as connectomes, represent estimates of the axonal-fiber pathways connecting the different regions. Connectomes considered in this application have been estimated as in \cite{DAI} via recently developed pipelines, which efficiently exploit structural MRI data to obtain a parcellation of the brain in anatomical regions, and dw-MRI images to recover the  fiber streamlines connecting each pair of brain regions; see \cite{DAI} for details. Posterior computation and inference is performed considering the same settings as in the application to creativity in \cite{DUR2}.
 
\section{Changes in Brain Network with Alzheimer's}
The global testing procedure in \eqref{eq:6} strongly favors the hypothesis of association between brain structural connectivity and AD diagnosis with $\hat{\mbox{Pr}}\{H_1 \mid    y_1, \mathcal{L}(A_1), \ldots, y_n,  \mathcal{L}(A_n)\} >0.99$. This confirms findings in \cite{DAI} highlighting significant  variations in brain network summary measures when comparing AD patients with cognitively healthy controls.

As expected the estimated significant differences between the edge probabilities in AD group and control group in Fig. 5 show an overall less connected brain network for the AD group compared to controls, in line with \cite{DAI}  and literature on AD. The main differences appear in terms of intra-hemispheric connections in the left hemisphere, while fewer local differences are found also in terms of inter-hemispheric connections and right intra-hemispheric. This major role of the left hemisphere agrees with \cite{DAI,THO}.
\begin{figure}[h]
\begin{center}
\centerline{\includegraphics[height=10cm, width=8.5cm]{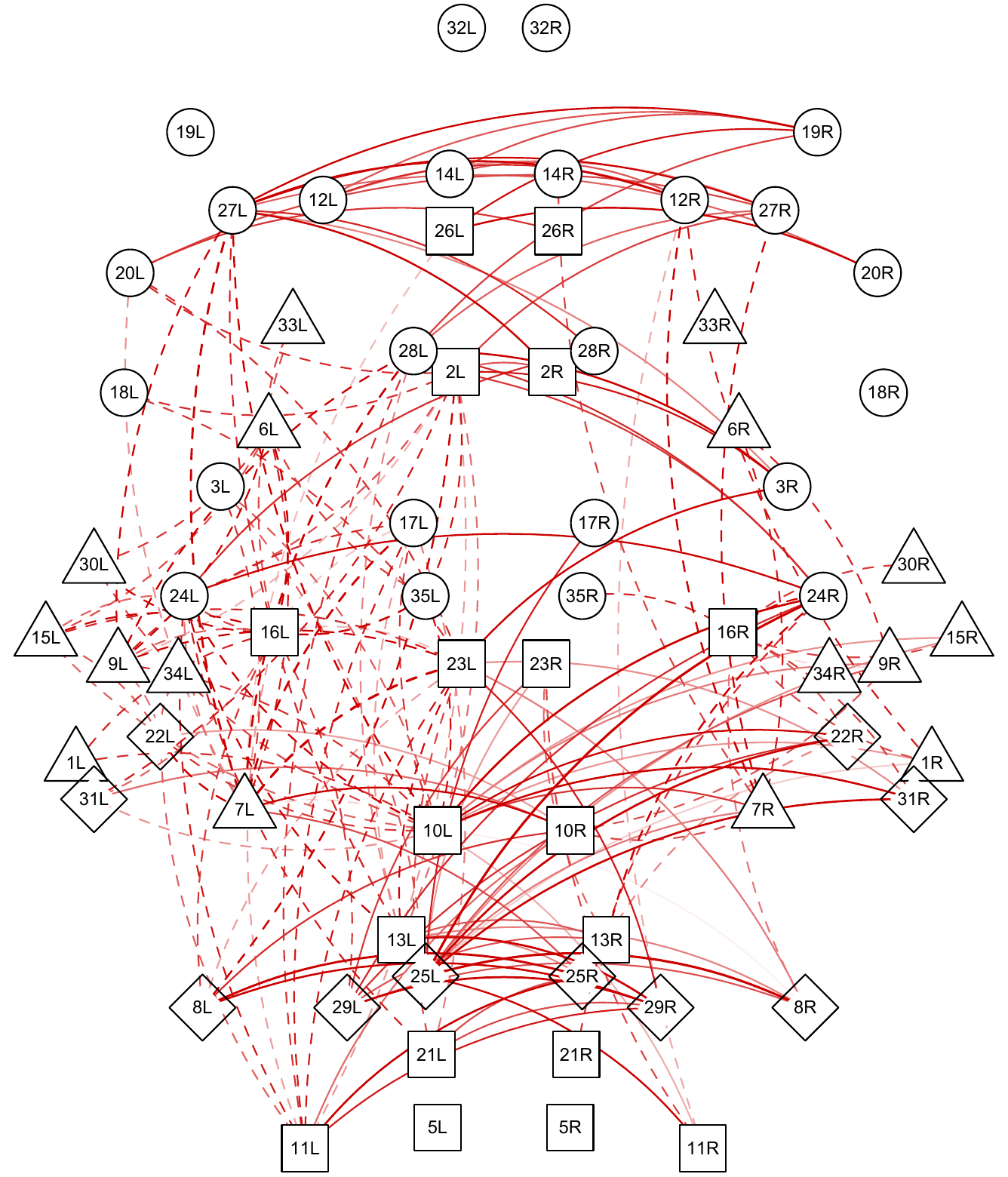}}
\caption{For connections significantly varying in AD according to \eqref{ro}, weighted network visualization with weights given by the estimated  $\mbox{Pr}\{\mathcal{L}(\mathcal{A})_l=1 \mid \mathcal{Y}=1\}-\mbox{Pr}\{\mathcal{L}(\mathcal{A})_l=1 \mid \mathcal{Y}=0\}=\sum_{h=1}^H \nu_{h1} \pi_l^{(h)}-\sum_{h=1}^H \nu_{h0} \pi_l^{(h)}$ $(l=1, \ldots, V(V-1)/2)$. Edges colors range from red to green as the corresponding difference goes from $-1$ to $1$. Solid lines refer to inter-hemispheric connections and dashed lines to intra-hemispheric connections. Frontal lobe regions (circles), Insular (ellipse), Limbic (square), Temporal (triangle), Parietal (diamond), Occipital (rectangle) according to classification of Desikan atlas in anatomical lobes \cite{KAN}.}
\label{F4}
\end{center}
\end{figure}

The agreement with previous studies highlights the consistency of our methodology, which has the additional benefit of providing inference not only on the scale of the network summary measures but in terms of variations of the entire pmf for the brain network-valued random variable representing brain interconnections. This rules out the issue of conflicting conclusions when different network statistics are considered, while also avoiding ad-hoc choices when defining certain summary measures. Recalling for example \cite{DAI}  one may obtain different results when considering an order for the k-core different from $18$. An additional benefit of our approach, as outlined in the simulation study, is that local testing automatically controls for multiplicity, while out-performing frequentist competitors controlling for FDR, in terms of power. Recalling the application to AD, this leads to a procedure which can more easily identify connections significantly varying between control and AD subjects. This is evident when comparing Fig. 5 to results in Fig. 1 in \cite{DAI} learning less significant local differences. This result may be related to the use of a region-specific network statistic which displays low variations across case and controls as well as the choice of an overly conservative level for the FDR and the less power related to mass-univariate local testing procedures. 

\begin{figure*}[t]
\centering
\includegraphics[height=5.2cm, width=18.5cm]{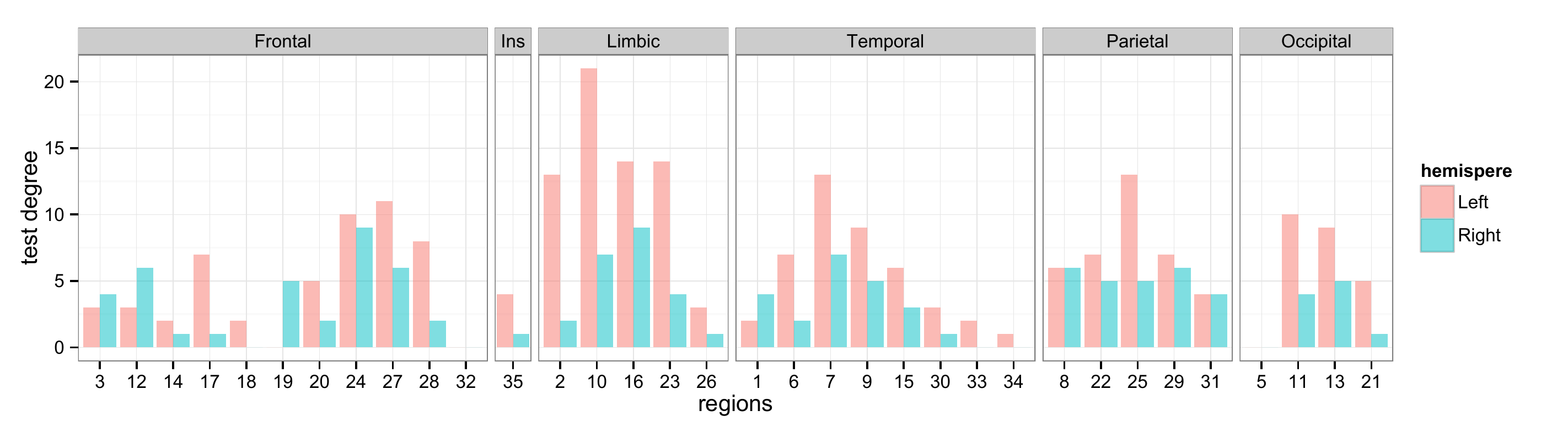}
\caption{{\small{Test degree for each brain region (classified in left and right hemisphere and corresponding lobe). The test degree of region $v$ is defined as the total number of connections among $v$ and the remaining $V-1$ regions significantly varying in Alzheimer's group.}}}
\label{F5}
\end{figure*}

Our approach doesn't rely on the choice of network summary measures and automatically controls for multiplicity, overcoming previous issues while strongly gaining power. As a result we learn more connections significantly varying between control and AD groups. This provides interesting new insights according to Fig. 6, which displays for each region $v$ $(v=1, \ldots, V)$ the total number  of connections among $v$ and the remaining $V-1$ regions significantly varying between controls and AD group under our local testing procedure \eqref{ro} with $\epsilon=0.1$. To highlight the roles of higher level brain systems, regions are grouped in  anatomical lobes according to  \cite{KAN} and in hemispheres.

Results in Fig. 6 highlight the connectivity breakdown for regions in the left hemisphere while providing new insights with respect to \cite{DAI}. In particular we learn the major role of regions in the left limbic lobe consistently with initial neuropathological studies \cite{HOP,BLE} and more recent empirical findings via MRI \cite{DEO,DESC} highlighting the key role of the limbic system in memory, attention and executive functioning, while focusing on this lobe as one of the areas mainly affected by AD. Significant changes are also found in the connectivity of the other anatomical lobes such as temporal, parietal and occipital, consistent with \cite{SMI,AZA,THO1,HOR}.

According to Fig. 6 the regions mostly affected by AD in terms of connectivity behavior are the left isthmus of the cingulate (10L), left parahippocampal (16L), left posterior cingulate (23L), left fusiform (7L) and left precuneus (25L). These results provide a unifying answer to different insights arising from several studies, typically focusing on the activity of a subset of regions. Parahippocampal atrophy is found in \cite{KES} and \cite{THA}; \cite{ZHO}   highlights abnormal connectivity in hippocampus and posterior cingulate, while \cite{KIM} learn reduced functional activity in hippocampus and precuneus, with the latter showing atrophy also in \cite{KAR}. Metabolic reduction in the posterior cingulate is studied in \cite{MIN} and \cite{LIA}. Reduced functional connectivity in the fusiform is found in \cite{GOL} and \cite{BOK} via fMRI. Fewer studies are available on the role of the isthmus of the cingulate with only a recent work of \cite{ZHU}  trying a first attempt in this direction. We provide a unifying vision, consistent with previous literature, while highlighting the  role of the isthmus. This region represents a bridge between the parahippocampal and the posterior cingulate, two critical regions extensively explored in the literature in terms of atrophy and metabolic reduction in AD subjects. Hence a reduced metabolic activity and increased atrophy of parahippocampal and the posterior cingulate, may be related to a disruption of the circuits from the left cingulate isthmus. 

Besides providing unifying novel results on brain network variations in AD, our methodology also represents the unique ability to assess evidence of AD according to the subject's full brain network structure. In fact, under our framework, the probability $\mbox{Pr}\{\mathcal{Y}=1 \mid \mathcal{L}(A_{i})\}$ that a subject $i$ has AD, conditionally on his brain's structural connectivity network $A_i$ is simply equal to
\begin{eqnarray}
\frac{p_{\mathcal{Y}}(1)p_{\mathcal{L}(\mathcal{A}) \mid 1}(a_i)}{p_{\mathcal{Y}}(1)p_{\mathcal{L}(\mathcal{A}) \mid 1}(a_i)+p_{\mathcal{Y}}(0)p_{\mathcal{L}(\mathcal{A}) \mid 0}(a_i)}, \label{eq:10}
\end{eqnarray}
where $\mathcal{L}(A_{i})=a_i$ is the network configuration of the $i$-th subject and $p_{\mathcal{L}(\mathcal{A}) \mid y}(a_i)$, $(y=0,1)$ can be easily computed from \eqref{eq:5}. Current classification procedures exploit either region activity vectors $\psi_i$ \cite{ESK} or network summary statistics vectors $\theta_i$ \cite{FRI,PRA}, rather than the whole brain network $\mathcal{L}(A_{i})$, to predict $y_i$. We evaluate our procedure in \eqref{eq:10} in terms of in-sample and out-of-sample classification performance. In the first case, we compute \eqref{eq:10} for each subject after considering all data in model estimation. Out-of-sample classification is instead performed by training the model on $69$ subjects and predicting the AD status via \eqref{eq:10} on the remaining one fourth of the individuals, with the training and test samples randomly selected. Our methodology provides an overall good classification performance, with an area under the ROC curve of $0.91$ for in-sample classification and $0.83$ for out-of-sample. The accuracy is instead $87\%$ in the former, and $75\%$ in the latter. These results out perform \cite{ESK}, and \cite{FRI} when summary statistics $\theta_i$ are extracted from undirected brain networks, while providing similar performance to \cite{PRA}. It is important to note that \cite{PRA} utilizes substantially more information in considering both weighted and flow connectivity networks for a total of $298{,}600$ network summary measures, rather than only binary connections encoding presence or absence of fibers. 

\section{Discussion}
Brain connectivity plays a key role in brain function and dysfunction. Modern magnetic resonance imaging technologies, combined with state-of-the-art data processing algorithms, have made it possible to reliably measure brain structural connectivity networks non-invasively. Understanding brain networks is necessary for developing improved diagnosis and treatment strategies for neurological disorders, but a deep understanding of the relationship between the network of structural interconnections in the brain and such disorders remains still elusive. 

The statistical methodology presented in this paper defines the first-ever probabilistic generative mechanism to draw tractable and efficient inference directly on the probability mass function associated to a network-valued random variable, rather than on network summary measures or multivariate activity data. In allowing the brain network data to be appropriately analyzed as network-valued, these methods enable substantial improvements in accurately detecting group differences, isolating specific aspects of the network that vary across neurological disorders, and enhancing performance of predictive models as outlined in the application to Alzheimer's disorder.

%----------------------------------------------------------------------------------------
%	REFERENCE LIST
%----------------------------------------------------------------------------------------
\phantomsection

%----------------------------------------------------------------------------------------


\begin{thebibliography}{10}
\bibitem{ASH}
Ashford~JW et al, eds (2011) {\em Handbook of Imaging the Alzheimer Brain} (IOS Press, Amsterdam).

\bibitem{EID}
Eidelberg~D, ed (2012) {\em Imaging In Parkinson?s Disease} (Oxford University Press, New York).

\bibitem{BUD}
Budson~AE, Kowall~NW (2011) {\em The Handbook of Alzheimer's Disease and Other Dementias} (Wiley-Blackwell, Oxford).

\bibitem{WHO}
World Health Organization (2006) {\em Neurological Disorders: Public Health Challenges} (WHO Press, Geneva).

\bibitem{FOD}
Fodor~JA (1983) {\em Modularity of Mind} (MIT Press, Cambridge).

\bibitem{CRA}
Craddock~RC et al (2013) Imaging human connectomes at
the macroscale. {\em Nature Methods} 10:524-539.


\bibitem{FUS}
Fuster~JM (2006) The cognit: a network model
of cortical representation. {\em Int J Psychophysiol} 60:125-132.

\bibitem{BRE}
Bressler~SL, Menon~V (2010) Large-scale brain networks in cognition: emerging methods and principles. {\em Trends Cogn Sci} 14:277-290.

\bibitem{DES}
Desikan~RS et al (2006) An automated labeling system for subdividing the human cerebral cortex on MRI scans into gyral based regions of interest. {\em Neuroimage} 31:968-980.

\bibitem{STA}
Stam~CJ (2014) Modern network science of neurological disorders. {\em Nat Rev Neurosci} 15:683-695.

\bibitem{KRZ}
Krzanowski~WJ, ed (1988) {\em Principles of multivariate analysis: a user's perspective} (Oxford University Press, New York).



\bibitem{RUB}
Rubinov~M, Sporns~O (2010) Complex network measures of brain connectivity: Uses and interpretations. {\em Neuroimage} 52:1059-1069.

\bibitem{ARD}
Arden~R, Chavez~RS, Grazioplene~R, Jung~RE (2010) Neuroimaging creativity: A
psychometric view. {\em Behav Brain Res} 214:143-156.

\bibitem{GEN}
Genovese~CR, Lazar~NA, Nichols~T (2002) Thresholding of Statistical Maps in Functional Neuroimaging Using the False Discovery Rate. {\em Neuroimage} 15:870-878.

\bibitem{FOR}
Fornito~A, Zalesky~A, Breakspear~M (2013) Graph analysis of the human connectome: Promise, progress, and pitfalls. {\em Neuroimage} 80:426-444.

\bibitem{BEN}
Benjamini~Y, Hochberg~Y (1995) Controlling the false discovery rate: A practical and powerful approach to multiple testing. {\em J R Stat Soc Series B Stat Methodol} 57:289-300.


\bibitem{ZAL}
 Zalesky~A, Fornito~A, Bullmore~ET (2010) Network-based statistic: Identifying differences in brain networks. {\em Neuroimage} 53:1197-1207.
 
\bibitem{HOFF}
Hoff~PD, Raftery~AE, Handcock~MS (2002) Latent space approaches to social network
analysis. {\em J Am Stat Assoc} 97:1090-1098.


\bibitem{DUR1}
Durante~D, Dunson~DB, Vogelstein~JT (2015) Nonparametric Bayes Modeling of
Populations of Networks. {\em arXiv:1406.7851}.

\bibitem{DUR2}
Durante~D, Dunson~DB (2015) Bayesian inference on group differences in brain networks. {\em arXiv:1411.6506}.

\bibitem{DUN}
Dunson~DB, Xing~C (2009) Nonparametric Bayes modeling of multivariate categorical
data. {\em J Am Stat Assoc} 104:1042-1051.

\bibitem{ROU}
Rousseau~J, Mengersen~K (2011) Asymptotic behaviour of the posterior distribution in
overfitted mixture models. {\em J R Stat Soc Series B Stat Methodol} 73:689-710.

\bibitem{BER}
Berger~JO, Selke~T (1987) Testing a point null hypothesis: The irreconcilability of p
values and evidence. {\em J Am Stat Assoc} 82:112-122.

\bibitem{AGR}
Agresti~A, ed (2002) {\em Categorical data analysis} (Second edition. New York: Wiley).

\bibitem{JAM}
James~BD et al (2014) Contribution of Alzheimer disease to mortality in the United States. {\em Neurology} 82:1-6.

\bibitem{HOP}
Hopper~M, Vogel~F (1976) The limbic system in alzheimer's disease. A neuropathologic investigation. {\em The American Journal of Pathology} 85:1-20.

\bibitem{THO}
Thompson~PM et al (2001) Cortical change in Alzheimer's disease detected with a disease-specific population-based brain atlas. {\em Cerabral Cortes} 11:1-16.

\bibitem{BOK}
Bokde~ALW et al (2006) Functional connectivity of the fusiform gyrus during a face-matching task in subjects with mild cognitive impairment. {\em Brain} 129:1113-1124.

\bibitem{DAI}
Daianu~M et al (2013) Breakdown of brain connectivity between normal aging and alzheimer's disease: A structural k-core network analysis. {\em Brain Connectivity} 3:407-422.

\bibitem{FISH}
Fischl~B et al (2004) Automatically parcellating the human cerebral cortex. {\em Cerebral Cortex} 14:11-22.

\bibitem{KAN}
Kang~X, Herron~TJ, Cate~AD, Yund~EW, Woods~DL (2012) Hemispherically unified surface maps of human cerebral cortex: Reliability and hemispheric asymmetries. {\em PLoS ONE} 9:e45582.

\bibitem{BLE}
Blesa~R, Mohr~E, Miletich~R, Chase~T (1995) Limbic system dysfunction in alzheimer's
disease. {\em Journal of Neurology, Neurosurgery \& Psychiatry} 59:450-451.

\bibitem{DEO}
Deoni~S et al (2011) Investigating limbic system myelin alteration in Alzheimer's disease. {\em Alzheimer's \& Dementia} 7:S57-S58.

\bibitem{DESC}
de Schotten~MT et al (2014) Advanced diffusion weighting imaging (DWI) tractography of the limbic system: novel biomarkers of neurodegenerative changes during progression/conversion from cognitive normality to ad dementia. {\em Alzheimer's \& Dementia} 10:P37.

\bibitem{SMI}
Smith~MZ, Esiri~MM, Barnetson~L, King~E, Nagy~ZS (2000) Constructional Apraxia in Alzheimer's Disease: Association with Occipital Lobe Pathology and Accelerated Cognitive Decline {\em Dementia and Geriatric Cognitive Disorders} 12:281-288.

\bibitem{AZA}
Azari~N et al (1992) Patterns of interregional correlations of cerebral glucose metabolic rates in patients with dementia of the Alzheimer's type. {\em Neurodegeneration} 1:101-111.

\bibitem{THO1}
Thompson~PM et al (2003) Dynamics of gray matter loss in Alzheimer's
disease. {\em The Journal of Neuroscience} 23:994-1005.

\bibitem{HOR}
Horwitz~B, Grady~C, Sclageter~N, Duara~R, Rapoport~S (1987) Intercorrelations of
regional glucose metabolic rates in Alzheimer's disease. {\em Brain Research} 407:294-306.

\bibitem{KES}
Kesslak~J, Nalcioglu~O, Cotman~C (1991) Quantification of magnetic resonance scans for hippocampal and parahippocampal atrophy in Alzheimer's disease. {\em Neurology} 41:51-54.

\bibitem{THA}
Thangavel~R, Hoesen~GV, Zaheer~A (2008) Posterior parahippocampal gyrus pathology in Alzheimer's disease. {\em Neuroscience} 154:667-676.

\bibitem{ZHO}
Zhou~Y et al (2008) Abnormal connectivity in the posterior cingulate and hippocampus in early Alzheimer's disease and mild cognitive impairment {\em Alzheimer's \& Dementia} 4:265-270.

\bibitem{KIM}
Kim~J, Kim~Y, Lee~J (2013) Hippocampusprecuneus functional connectivity as an early sign of Alzheimer's disease: A preliminary study using structural and functional magnetic resonance imaging data. {\em Brain Research} 23:18-29.

\bibitem{KAR}
Karas~G et al (2007) Precuneus atrophy in early-onset Alzheimer's disease: a
morphometric structural MRI study. {\em Neuroradiology} 49:967-976.

\bibitem{MIN}
Minoshima~S et al (1997) Metabolic reduction in the posterior cingulate cortex in very early alzheimer's disease. {\em Annals of Neurology} 42:85-94.

\bibitem{LIA}
Liang~WS et al (2008) Alzheimer's disease is associated with reduced expression
of energy metabolism genes in posterior cingulate neurons. {\em Proceedings of the National Academy of Sciences} 105:4441-4446.

\bibitem{GOL}
Golby~A et al (2005) Memory encoding in alzheimer's disease: an fmri study of explicit and implicit memory. {\em Brain} 128:773-787.



\bibitem{ZHU}
Zhu~D, Majumdar~S, Korolev~I, Berger~K, Bozoki~A (2013) Alzheimer's disease and amnestic mild cognitive impairment weaken connections within the default-mode network: a multi-modal imaging study. {\em Journal of Alzheimer's Disease} 34:969-984.

\bibitem{ESK}
Eskildsen~SF, Coup\'e~P, Fonov~VS, Pruessner~JC, Collins~DL (2015)  Structural imaging biomarkers of Alzheimer's disease: predicting disease progression. {\em Neurobiology of Aging} 36:S23-S31.

\bibitem{FRI}
Friedman~EJ et al (2014) Directed Progression Brain Networks in Alzheimer's Disease: Properties and Classification. {\em Brain Connectivity} 5:384-393.

\bibitem{PRA}
Prasad~G, Joshi~SH, Nir~TM, Toga~AW, Thompson~PM (2014) A Brain connectivity and novel network measures for Alzheimer's disease classification. {\em Neurobiology of Aging} 36:S121-S131.




\end{thebibliography}
\end{document}